\documentclass{mnras}

\usepackage{graphicx}
\usepackage{subfigure}
\usepackage{txfonts}
\let\bibhang\relax
\usepackage{natbib}
\bibpunct{(}{)}{;}{a}{}{,}

\title[]{Multi-wavelength, spatially resolved modelling of {HD~48682's} debris disc}

\author[S. Hengst et. al.]{S. Hengst$^{1}$\thanks{Contact e-mail: \href{mailto:U1082091@umail.usq.edu.au}{shane.hengst@usq.edu.au}}, {J. P. Marshall$^{2,1} $}, \ {J. Horner$^{1}$},\  and {S. C. Marsden$^{1}$}\\
$^{1}$Centre for Astrophysics, University of Southern Queensland, Toowoomba, QLD 4350, Australia\\
$^{2}$Academia Sinica Institute of Astronomy and Astrophysics, 11F of AS/NTU Astronomy-Mathematics Building,\\No.1, Sect. 4, Roosevelt Rd, Taipei 10617, Taiwan\\}

\date{Last updated ---; in original form ---}

\pubyear{2020}

\begin{document}
\label{firstpage}
\pagerange{\pageref{firstpage}--\pageref{lastpage}}
\maketitle

\begin{abstract}
Asteroids and comets (planetesimals) are created in gas- and dust-rich protoplanetary discs. The presence of these planetesimals around main-sequence stars is usually inferred from the detection of excess continuum emission at infrared wavelengths from dust grains produced by destructive processes within these discs. Modelling of the disc structure and dust grain properties for those discs is often hindered by the absence of any meaningful constraint on the location and spatial extent of the disc.  Multi-wavelength, spatially resolved imaging is thus invaluable in refining the interpretation of these systems. Observations of HD~48682 at far-infrared (\textit{Spitzer}, \textit{Herschel}) and sub-millimetre (JCMT, SMA) wavelengths indicated the presence of an extended, cold debris disc with a blackbody temperature of 57.9~$\pm$~0.7~K. Here, we combined these data to perform a comprehensive study of the disc architecture and its implications for the dust grain properties. The deconvolved images revealed a cold debris belt, verified by combining a 3D radiative transfer dust continuum model with image analysis to replicate the structure using a single, axisymmetric annulus. A Markov chain Monte Carlo analysis calculated the maximum likelihood of HD48682's disc radius ($R_{disc} = 89^{+17}_{-20}~$au), fractional width ($\Delta R_{\rm disc} = 0.41^{+0.27}_{-0.20}$), position angle ($\theta = 66\fdg3^{+4.5}_{-4.9}$), and inclination ($\phi = 112\fdg5^{+4.2}_{-4.2}$). HD~48682 has been revealed to host a collisionally active, broad disc whose emission is dominated by small dust grains, $s_{\rm min} \sim 0.6~\mu$m, and a size distribution exponent of $3.60~\pm~0.02$.

\end{abstract}

\begin{keywords}
stars: individual: HD~48682 -- circumstellar matter -- infrared: stars: planetary systems
\newline
\newline
\textbf{Accepted to appear in the Monthly Notices of the Royal Astronomical Society on 20th June, 2020}
\end{keywords}



\section{Introduction}

Planets and planetesimals (asteroids and comets) are known to be forged during the earliest stages of stellar evolution in dusty, gaseous circumstellar material known as a protoplanetary disc \citep{1993Lissauer,2015Wyatt}.  Once the primordial protoplanetary disc has dissipated and planet formation processes have (mostly) run their course, planetesimal belts are still observable around stars through the detection of scattered light and continuum emission at infrared and millimetre wavelengths from dust produced in collisions between planetesimals. Such systems are therefore referred to as `debris discs'; their study forms an integral part of our understanding of the formation and evolution of planetary systems \citep{2008Wyatt,2014Matthews,2018AHughes}.  

Initial studies into debris discs were mostly limited to the investigation of spatially unresolved photometry \citep[e.g. ][]{1993Backman,2007Wyatt,2011Kains} and/or mid-infrared spectroscopy \citep{2006Chen,2006aBeichman, 2014Chen, 2015Mittal}. Such data only provided weak constraints on the location and spatial extent of the debris discs. More recently, spatially resolved images particularly from \textit{Herschel} and ALMA, but also scattered light facilites such as \textit{Hubble}, GPI, and SPHERE, have revealed the underlying structure within a greater number of discs \citep[e.g.][]{2013Booth,2014Pawellek,2015Moor,2016Morales,2017Holland,2018Matra}. Observations at different wavelengths are sensitive to dust grains with different sizes and temperatures. Scattered light, and near- and mid-infrared wavelengths are sensitive to smaller dust grains whose orbital motions and spatial distribution are influenced by radiation forces; whereas millimetre wavelengths trace larger (cooler) grains whose orbital motions are more weakly affected by radiation forces \citep{1979BurnsLamySoter}. Multi-wavelength, spatially resolved imaging therefore provides better insight into both the location of planetesimal belts, and the properties of the dust grains generated by those bodies. The combination of imaging and photometric information addresses inherent degeneracies in the modelling of either data set individually \citep[e.g.][]{2014bMarshall}, enabling the most detailed possible understanding of these systems' properties \citep[e.g.][]{2012Ertel,2012Lohne,2014Ertel,2017Hengst,2019Geiler}.

Debris discs have been detected around hundreds of stars in either scattered light or continuum emission, but a substantial fraction of these systems remain spatially unresolved \citep[e.g.][]{2014Matthews,2018AHughes}.  However, a survey of 34 resolved discs conducted in the far-infrared wavelengths determined relationships between dust grain sizes, minimum grain size ($s_{min}$) and radiation blowout size ($s_{blow}$), with stellar luminosities and dust temperatures \citep{2014Pawellek}. These relationships can be applied to understanding unresolved systems. For example, by exploring the ratio of $s_{min}/s_{blow}$ to obtain insights into the nature of dusty grains across a range of environments, and infer the expected extent of those spatially unresolved discs \citep{2015PawKri}. 

An increasing number of debris discs have been spatially resolved at millimetre wavelengths in recent years, particularly fuelled by the increased capabilities of ALMA but previously driven by observations by the JCMT and SMA \citep[e.g.][]{2016Steele, 2017Holland,2018Matra,2019Sepulveda}. \cite{2018Matra} determined a strong statistical relationship between the stellar luminosity and the measured planetesimal belt radii based on spatially resolved observations of 26 debris discs with the SMA and ALMA. \cite{2018Matra} suggested there may be a link between CO snow lines in protoplanetary discs and the subsequent locations of planetesimal belts \citep[e.g.][]{2018Andrews,2020Pinte}. 

HD~48682 (56~Aur; HIP~32480) was originally classified as a visual binary in the Washington double-star catalogue \citep{2001Mason}.  The star was first identified to have extended excess by the \textit{InfraRed Astronomical Satellite} \citep[\textit{IRAS},][]{1991Aumann}, although there was confusion as to which star in the system was responsible for the extended emission.  It was later discovered that the two stars were not physically associated with each other due to differing proper motions, and \cite{2004Sheret} concluded that the large 60 $\micron$ \textit{IRAS} excess is associated with the `primary' as confirmed from their Submillimetre Common-User Bolometer Array (SCUBA) observation.  The primary is referred to as HD~48682 and is classified as G0 V star \citep{1995Hoffleit}, whilst the secondary is referred to as HD~48682B, is classified as a M0 star and is not a component of this work.  HD~48682 was spatially resolved using the Multiband Infrared Photometer (MIPS) camera \cite{2004Rieke} onboard the \textit{Spitzer} Space Telescope \citep{2004Werner} at 70 $\micron$ \citep{2005Stapelfeldt}. 

Extensive archival imaging data sets of HD~48682 are available in the far-infrared and sub-millimetre wavebands.  It was observed as part of the Open Time Key Programme Dust around Nearby Stars \citep{2013Eiroa} with the \textit{Herschel} Space Observatory \citep{2010Pilbratt} in six wavebands from 70 to 500~$\micron$ with the Photodetector Array Camera and Spectrometer and Spectrometer and the Photometric Imaging REceiver instruments \citep[PACS and SPIRE;][]{2010Poglitsch,2010Griffin}.  The {Sub-millimetre} Common-User Bolometer Array 2 (SCUBA-2) on the James Clerk Maxwell Telescope \citep[JCMT,][]{2013Holland} produced images in the wavebands 450 and 850 $\micron$ for HD~48682. The Sub-Millimeter Array (SMA), an 8-element interferometer that covers between 180 to 800 GHz \citep{2004Ho}, obtained data in the vicinity of HD~48682 with the receiver band centred on 225 GHz ($\approx$~1.3~mm).

In this work, we analyse the available far-infrared and sub-millimetre images of HD~48682 in combination with archival photometric and the mid-infrared spectroscopic data, to seek a better understanding of the architecture and composition of HD~48682's debris disc. In Section \ref{s:OA}, we describe the observations, the associated modelling and analysis of HD~48682.  In Section \ref{s:RD}, the results of the analysis of the Spectral Energy Distribution (SED), images and radial profiles at \textit{Herschel}/PACS 70/100/160 $\micron$ are presented along with calculations of the disc's observational properties: dust temperature $T_{\rm dust}$, disc radial extent $R_{\rm dust}$ (both from the blackbody assumption and the resolved/deconvolved images), and disc fractional luminosity $L_{\rm dust}/L_{\star}$. We discuss the fitting of the imaging and photometric observations using the 3D MCMC radiative transfer code {\sc{Hyperion}} \citep{2011Robitaille} and the determination of most probable disc parameters and associated uncertainties using \textsc{emcee} \citep{2013Foreman-Mackey}. In Section \ref{s:D}, we discuss the state of the disc in comparison with other studied systems and the potential disc brightness asymmetry observation. Finally, in Section \ref{s:C}, we present our conclusions.

Whilst these research procedures have been used to study various circumstellar discs, they have not been employed in this combination specifically for debris discs. We view this as a first paper of a series to provide a foundation in the analysis of debris discs.  This paper will be followed by \citet[\textit{in review}]{2020Marshall} that will employ these procedures for a number of debris disc systems.

\section{Observations and analysis}\label{s:OA}

The observational data for the HD~48682 system are presented in this section.  This includes describing the characterisation of the host star through fitting a stellar photosphere model, a summary of the ancillary photometry compiled for the SED, and a description of the reduction of the \textit{Herschel}/PACS observations and subsequent analysis. The compiled SED for HD~48682, the scaled photospheric model, and the blackbody fit to the dust excess are shown in Figure \ref{fig:SED}.

\subsection{Stellar parameters}

HD~48682 is a nearby \cite[\textit{d} = 16.65 $\pm$ 0.07 pc;][]{2018G2} G0 main-sequence star with an effective temperature 6086 K \citep{2013Eiroa}. Several attempts have been made to determine the age of HD~48682, yielding a wide range of results.  Most recently, \cite{2013Eiroa} obtained two distinct estimates of the age of HD~48682 through the use of X-ray observations and Ca~II as tracers.  These two methods yielded ages of 1.38 and 6.38~Gyr respectively. Where isochrones were used to determine stellar age, \cite{1977Perrin}, from a sample of 138 stars, and \cite{2009Holmberg}, from a sample of 16 682 F and G stars,  calculated ages of 8.91 Gyr and 3.2$^{+1.4}_{-1.9}$ Gyr respectively.  These calculated stellar ages are used with caution as \cite{2009Holmberg} states that isochrones are very sensitive to the star’s effective temperature and metallicity.  However, all methods agree that HD~48682 stellar age is at least 1 Gyr. A summary of the stellar physical properties for HD~48682 is given in Table \ref{tab:StarParam}.

\begin{table}
 \caption{Physical properties of HD~48682.  \label{tab:StarParam}}
 \centering 
 \begin{tabular}{lll}
  \hline
  Parameter & Value & Ref.\\
  \hline\hline
  Distance [pc] & 16.65 $\pm$ 0.07 & 1\\
  Right Ascension [h:m:s] (J2015.5) & $06:46:44.33$ & 1 \\
  Declination [d:m:s] (J2015.5) & $+43:34:41.28$ & 1 \\
  Proper Motions (RA,Dec) [mas/yr] &  -3.127 $\pm$ 0.341 & 1 \\ 
    \                           & 63.583 $\pm$ 0.337 & 1 \\
  Spectral \& Luminosity Class & G0 V & 2 \\
  $V$[mag] & 5.200 $\pm$ 0.031 & 3\\
  $B$ - $V$[mag] & 0.573 $\pm$ 0.015& 3\\
  Bolometric Luminosity [$L_\odot$] & 1.752 & 2\\
  Mass [$M_{\odot}$] & 1.17 $\pm$ 0.04 & 2 \\
  Temperature [K] & 6086 $\pm$ 50, 6054$^{+65}_{-39}$ & 2, 1 \\
  Surface Gravity, $\log g$ [cm/s$^{2}$] & 4.35 $\pm$ 0.03, 4.5 & 2, 1 \\
  Metallicity [Fe/H] & 0.09 $\pm$ 0.09 &  2\\
  Age [Gyr] & 1.38 (X-ray) & 2\\
  & 6.32 (Ca II) & 2\\
   & 3.2$^{+1.4}_{-1.9}$ & 4 \\
    & 8.91 & 5 \\
  Radius [$R_{\odot}$]& 1.18$^{+0.14}_{-0.02}$ & 6\\
  \hline
 \end{tabular}

\medskip
\raggedright
 {References.} (1) \cite{2018G2};  (2) \cite{2013Eiroa}; (3) \cite{1993Turon}; (4) \cite{2009HolmCat}; (5) \cite{1977Perrin}; (6) \cite{2007Takeda}.
\end{table}

The stellar photospheric contribution to HD~48682's SED was modelled using the closest matching model ($T_{\rm{eff}}$ = 6030 K, log~$g$ = +4.39, [Fe/H] = 0.0) available from the Castelli-Kurcz atlas\footnote{Castelli-Kurcz models can be obtained from: http://www.stsci.edu/hst/observatory/crds/castelli\_kurucz\_atlas.html} \citep{2004CK}. The selected photospheric model was scaled to the optical and infrared observations at wavelengths between 0.4 and 10 $\micron$, weighted by their uncertainties, using a least squares fit ($\chi^2$~=~5.28,~$\chi^2_{red}$~=~1.06).  HD~48682's assumed stellar radius from this scaling of the photospheric model was determined to be 1.23 $R_\odot$, consistent with the estimate of 1.18$^{+0.14}_{-0.02}~R_\odot$ calculated by \cite{2007Takeda}.

\subsection{Ancillary data}

\begin{table}
\caption{Photometry of HD~48682. \label{tab:Photometry}}
\centering
\begin{tabular}{lrr}
\hline
Wavelength  & Flux & Reference \\
  $[\micron]$ & [Jy] & \\
\hline\hline
0.440 & 19.62 $\pm$ 0.84 & 1 \\
0.550 & 28.67 $\pm$ 0.24 & 1 \\
0.71 & 34.30 $\pm$ 1.39 & 2\\
1.25 & 31.699 $\pm$ 1.494 & 3\\
1.65 & 26.851 $\pm$ 1.266 & 3 \\
2.20 & 17.278 $\pm$ 0.868 & 3 \\
3.40 & 8.666 $\pm$ 2.727 & 4 \\
4.60 & 4.5114 $\pm$ 1.132 & 4\\
9 & 1.3534 $\pm$  0.0451 & 5 \\
12  &  0.761 $\pm$ 0.011 &  4 \\
18   &  0.4625 $\pm$ 0.0631 & 5 \\
22   &  0.244 $\pm$ 0.005 & 4 \\
30 & 0.148 $\pm$  0.013  & 6\\
32 & 0.142 $\pm$  0.017  & 6\\
34 & 0.136 $\pm$  0.020   & 6\\
70   &  0.290  $\pm$ 0.038 & 7 \\
70   &  0.264  $\pm$ 0.004 & 8 \\
100  &  0.275 $\pm$ 0.007 & 7 \\
100   &  0.252  $\pm$ 0.003 & 8  \\
160  &  0.177 $\pm$ 0.024 & 7 \\
160   &  0.182  $\pm$ 0.005 & 8 \\
250 & 0.090 $\pm$ 0.015 & 8\\
350 & 0.025 $\pm$ 0.008 & 8\\
450 & $<$ 0.025 & 9 \\
500 & $<$ 0.024 & 8\\
850 & 0.0039 $\pm$ 0.0008 & 9\\
\hline
\end{tabular}

\medskip
\raggedright
\textbf{References.} (1) \textit{Hipparcos} Input catalogue V2, \cite{1993Turon}; (2) SDSS/Johnson-Cousins, \cite{2008Just}; (3) Johnson-UKIRT, \cite{1999Gezari}; (4) \textit{WISE} all-sky survey, \cite{2010Wright}; (5) \textit{AKARI} IRC all-sky survey, \cite{2010Ishihara}; (6) \textit{Spitzer}/IRS synthetic photometry, this work; (7) \textit{Herschel}/PACS, this work; (8) \textit{Herschel}/PACS+SPIRE, \cite{2013Eiroa}; (9) JCMT/SCUBA-2 \cite{2017Holland}. 
\newline
\end{table}

To model the SED of HD~48682, the \textit{Herschel} and SCUBA-2 photometry were supplemented with a broad range of observations from the literature spanning optical to far-infrared wavelengths.  A summary of the collected photometry is shown in Table \ref{tab:Photometry}.  

The optical Johnson \textit{BV} photometry were taken from the \textit{Hipparcos} Input Catalogue \citep{1993Turon}, whilst the near-infrared Cousins \textit{I} and Johnson \textit{JHK$_s$} photometry were taken from the Sloan Digital Sky Survey \citep[SDSS;][]{2008Just} and United Kingdom Infra-Red Telescope \citep[UKIRT;][]{1999Gezari}, respectively. We opted not to use the 2MASS photometry because of lower precision and note that the 2MASS images were saturated by the secondary star \citep{2004Sheret}.

\begin{figure}
\includegraphics[width=\columnwidth, trim = 1cm 1cm 1cm 1cm ]{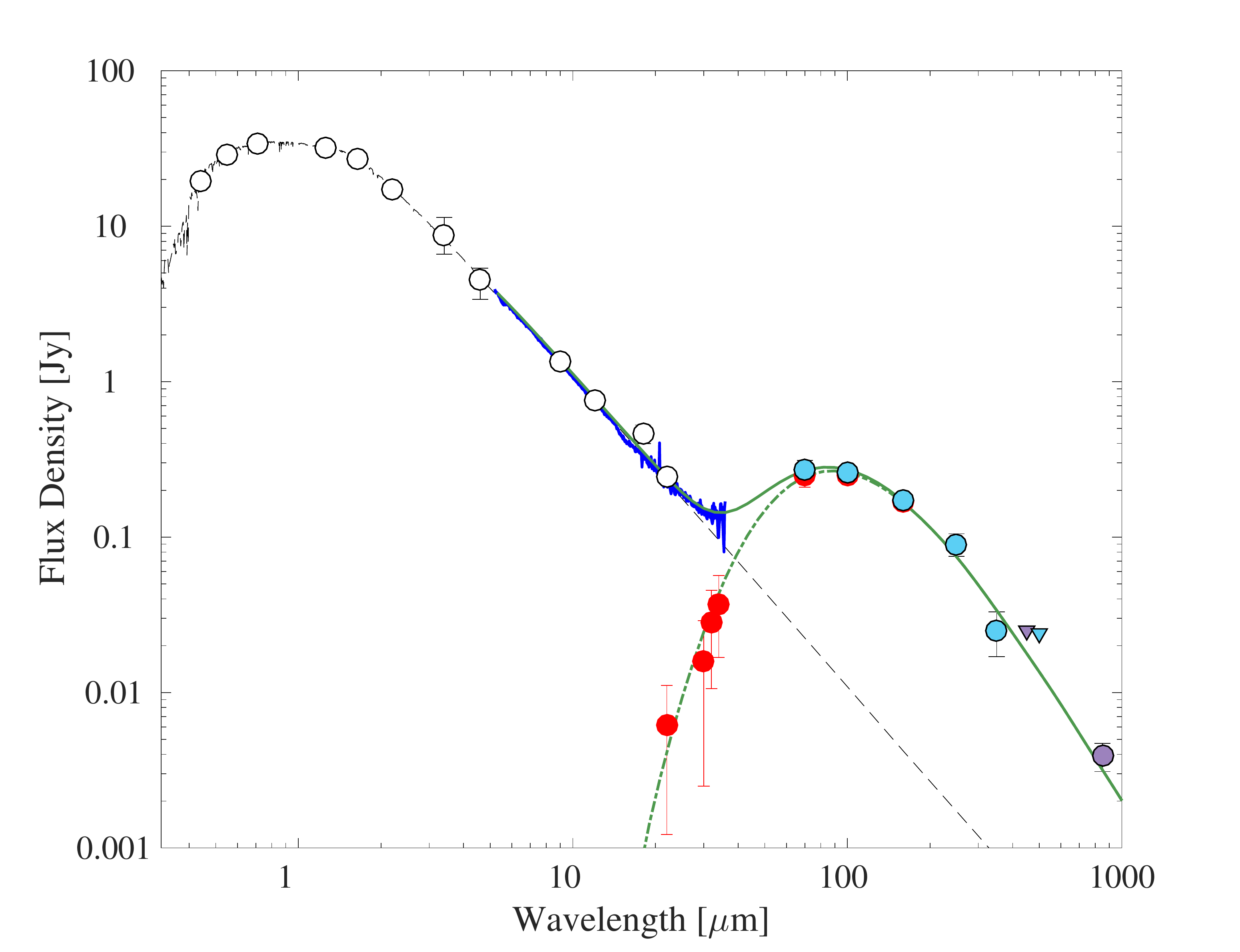}
\caption{SED of HD~48682. \label{fig:SED}
Circle data points are the measured photometry for HD~48682 ranging from optical to sub-millimetre wavelengths. Photometry between 0.4 to 10 $\micron$ were used to scale the stellar photosphere model represented by the (black) dashed line. The broad (blue) solid line is the \textit{Spitzer} IRS spectrum. The light blue data points represent the \textit{Herschel} photometry.  The purple data points represent the SCUBA-2 photometry. Solid triangle data points show the upper limit of 450 $\micron$ and 500 $\micron$ flux values. Solid Circle (red) data points represent the stellar subtracted values to calculate the (cold) dust contribution, estimated as a 57.9 K blackbody, which is represented as a (green) dotted line. The total flux and disc-only contribution are shown as solid and dot-dashed green lines, respectively. See Section \ref{s:rt} for the further details.}
\end{figure}

The mid-infrared photometry that were used were taken from the \textit{AKARI} IRC all-sky survey at 9 and 18 $\micron$ \citep{2010Ishihara} and the \textit{WISE} survey at 3.4, 4.6, 12, and 22 $\micron$ \citep{2010Wright}.  Colour corrections were applied to the \textit{AKARI} IRC measurements assuming a blackbody temperature of 6000 K (factors of 1.180 at 9 $\micron$ and 0.990 at 18 $\micron$) and applied to the \textit{WISE} photometry assuming a Rayleigh-Jeans slope.  A correction (factor of 0.873) was applied before the flux conversion to the \textit{WISE} 4.6 $\micron$ catalog (magnitude) value because of a known photometric bias \citep{2018Tiss}.  It can be seen that the \textit{AKARI} IRC 18 $\micron$ measurement has a slight excess, which could be attributed to either a warm component of the debris disc (although this was ruled out by \citealt{2014Pawellek}), the secondary star nearby (although the 9 $\micron$ measurement appears to be unaffected), or, given the wavelength range (13.9 - 25.6 $\micron$) of the corresponding filter, the initial rise of the cold excess emission from $\sim 20~\micron$ (see Figure \ref{fig:SED}).

A \textit{Spitzer} InfraRed Spectrograph \citep[IRS;][]{2004Houck} low-resolution spectrum spanning $\sim$5 to 36 $\micron$ was taken from the CASSIS database\footnote{The Cornell Atlas of \textit{Spitzer} IRS Sources (CASSIS) is a product of the Infrared Science Center at Cornell University, supported by NASA and JPL.} \citep{2011Lebouteiller}.  The IRS spectrum was scaled by the weighted mean differences at wavelengths $< 10 \micron$, where significant excess from the debris disc is not expected. The IRS spectrum was scaled using synthetic photometry extracted from the spectrum using the \textit{AKARI} IRC9 and \textit{WISE} W3 filter passbands, from which we determined a best-fit scaling factor of 0.94. For it to be included in the SED modelling process, the IRS spectrum was binned with a weighted average mean (with the associated uncertainty) to calculate synthetic photometry at 30, 32, and 34 $\micron$.  These values were used to trace the rise in the excess emission from the dust above the photosphere at mid-infrared wavelengths (see Section \ref{s:DiscArc} for details). 

\subsection{\textit{Herschel} data}

HD~48682 was observed as part of the \textit{Herschel} Open Time Key Programme DUNES \citep[DUst around NEarby Stars; KPOT\_ceiroa\_1)][]{2013Eiroa}.  PACS scan map observations of HD~48682 were taken in both 70/160 $\micron$ and 100/160 $\micron$ channel combinations.  SPIRE observations were taken in the 350/450/500 channel combination. The log of \textit{Herschel} observations is given in Table \ref{tab:obsers}. 

\begin{table}
 \caption{Summary of \textit{Herschel} PACS and SPIRE observations of HD~48682. \label{tab:obsers}}
 \centering
 \begin{tabular}{llll}
  \hline
  Instrument & Observation ID & Wavelengths [$\micron$] & Observation Date\\
  \hline\hline
PACS & 1342219021/22 & 70/160 & 19-Apr-2011\\
PACS & 1342206334/35 & 100/160 & 12-Oct-2010 \\
SPIRE & 1342204066 & 350/450/500 & 5-Sep-2010\\
  \hline
 \end{tabular}
 \newline
\end{table}

The Level 3 image products of the \textit{Herschel}/PACS data were directly downloaded from the Herschel Science Archive\footnote{http://archives.esac.esa.int/hsa/whsa/}. The image scales for these final mosaicked images were 1\farcs6 per pixel for the 70 and 100 $\micron$ images, and 3\farcs2 per pixel for the 160 $\micron$ image. 

Fluxes were measured using a standard Aperture Photometry routine implemented in the Herschel Interactive Processing Environment\footnote{{\sc hipe} is a joint development by the Herschel Science Ground Segment Consortium, composed of ESA, the NASA Herschel Science Center, and the HIFI, PACS and SPIRE consortia. http://www.cosmos.esa.int/web/herschel/hipe-download} \cite[{\sc hipe} - user release 15.0.1 and PACS calibration version 78 - the latest available public release at the time; ][]{2010Ott} . The circular aperture radii were chosen to be 12 pix (19\farcs2), 11 pix (17\farcs6), and 9 pix (28\farcs8) for the 70, 100, and 160 $\micron$ maps respectively. To estimate the value of the sky noise background we first measured the median values of ten randomly placed 25x25 pixel sub-regions for 70 and 100 $\micron$ images and 10x10 sub-regions for 160 $\micron$ image to avoid HD~48682 and borders pixel values - similar to the method used in \cite{2017Hengst}.  To determine a final sky noise estimate for each image we adopted the method outlined in \cite{2013Eiroa}. The mean of median sub-region values was used as the estimate of the dispersion of background flux ($\sigma_{pix}$). The sky noise was calculated by multiplying $\sigma_{pix}$ by the product of the square root of the number of pixels ($\sqrt{N^{circ}_{pix}}$) used in the respective circular apertures in each image for the flux estimate (Poisson noise) and the correlated noise component ($\alpha_{corr}$ = 3.7) as determined by \cite{2013Eiroa}. To yield final photometric values, the calculated sky noise values were subtracted from the flux estimates measured in each circular aperture to yield the source flux of HD~48682 for PACS images.  

The source flux values were scaled by appropriate aperture correction factors of 0.859 (19\farcs2), 0.835 (17\farcs6) and 0.849 (29\farcs8) at 70, 100 and 160 $\micron$ photometry, respectively \citep{2014Balog}. It is noted that it may be contentious when applying aperture corrections calculated for point sources when working on extended sources, however, this method has been adopted widely in the literature \citep[e.g.][]{2017Hengst}. The flux uncertainties were estimated from the standard deviation of the median values of the sub-regions respectively for each image, and then multiplied by the Poisson noise distribution and the correlated noise value as before. The final PACS fluxes and associated uncertainties are shown in Table \ref{tab:Photometry}, alongside the PACS estimates from \cite{2013Eiroa} for comparison.
 
It was discovered that the deconvolution of the 70 $\micron$ map revealed background or noise sources beyond the NW and SE arms of the disc that may contribute to confusion noise (see Section \ref{bgs} for details).  As the circular aperture used in the photometry image would cover these sources, then the flux estimate of these sources ($\approx$ 0.03 Jy) was used to calculate the final uncertainty corresponding to the 70 $\micron$ flux.

\section{Results}\label{s:RD}

In this section, we present the results of our analysis of HD~48682's disc. We first examine the \textit{Herschel}/PACS images for evidence of extended emission, followed by the application of a deconvolution routine to ascertain the disc extent and geometry. The additional photometric points were combined with the ancillary data and the measured radial extent from the images to model the disc, by fitting the excess emission with a modified blackbody model, employing an invariate Markov Chain Monte Carlo analysis, and applying a 3D continuum Gaussian model from which we deduce the disc structure and dust grain properties. We also present SCUBA-2 and SMA sub-millimetre image data pertaining to HD~48682.

\subsection{Resolved Images}\label{RD:DiscImages}

\begin{figure*}
\centering
\includegraphics[width=2\columnwidth,trim = 2cm 0cm 2cm 0cm]{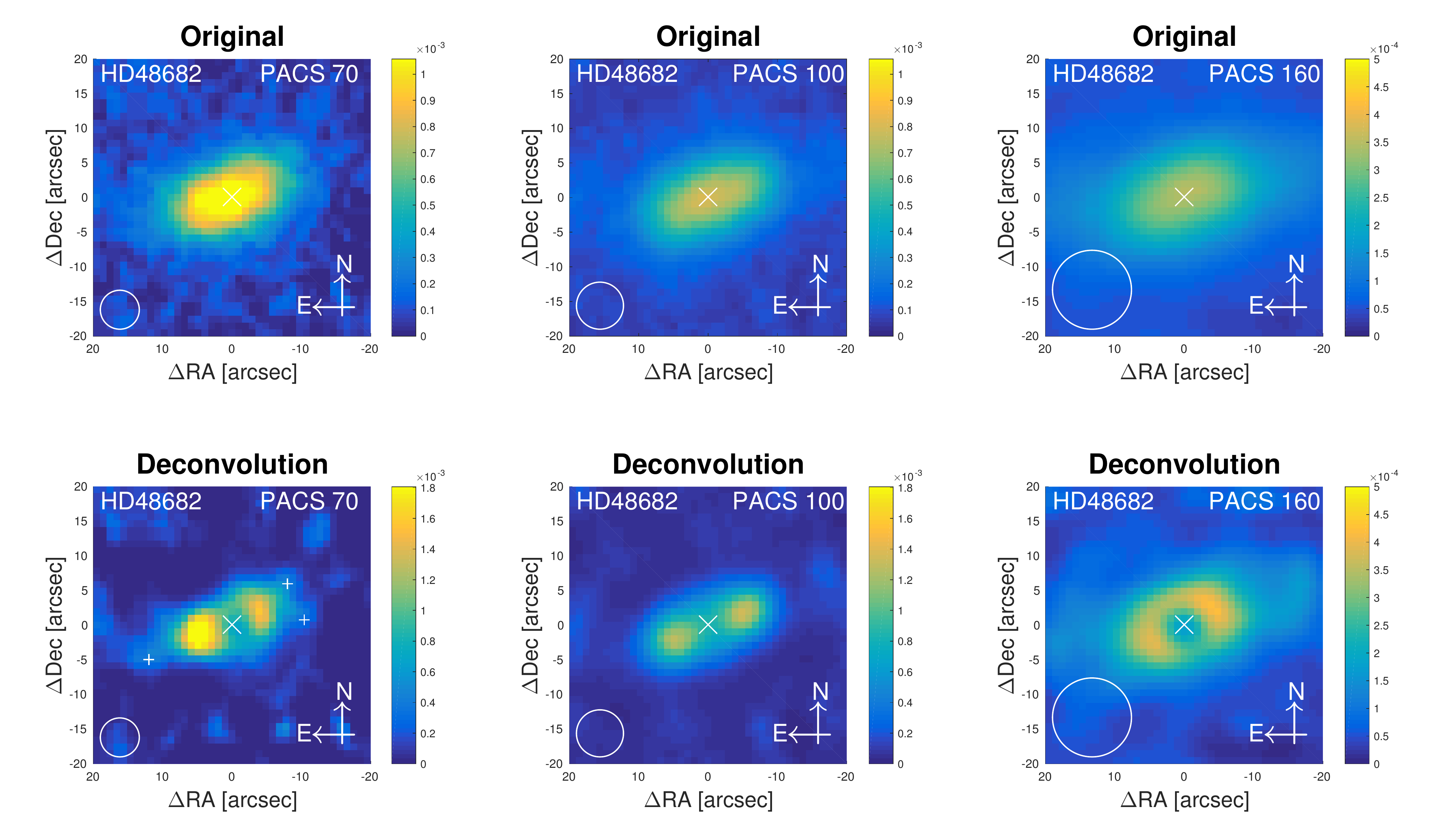}
\caption{Original \textit{Herschel} resolved image maps of HD~48682 (top row), the Lucy-Richardson deconvolved images of the original maps after star subtraction (bottom row). The star's position is represented by a white cross. The instrument beam size (FWHM) is represented by the circle in the bottom left-hand corner for each image. The additional sources, which could be regarded as an extension of the disc, background contamination, or noise peaks in the 70 $\micron$ deconvolution, are represented by white plus-signs. Orientation of the image is north up, east left. The images have been re-scaled to $1\arcsec$ pixel$^{-1}$.  The colour scale bar is in units of Jy/arcsec$^{2}$. The image RMS (in Jy/arcsec$^{2}$) values are $2.9\times10^{-4}$ at 70 $\micron$, $2.3\times10^{-4}$ at 100 $\micron$, and $1.3\times10^{-4}$ at 160 $\micron$. 
\label{fig:stamps}}
\end{figure*}

HD~48682 was resolved in the far-infrared wavebands by the \textit{Herschel}/PACS instrument \citep{2010Poglitsch} showing the extent of the disc for wavelengths 70, 100, and 160 $\micron$ (see top row in Figure \ref{fig:stamps}). The 70 and 100 $\micron$ maps were originally 0.6250 pixels per 1\arcsec, whilst the 160 $\micron$ map was 0.3125 pixels to 1\arcsec.  All maps were re-sampled to 1 pixel to 1\arcsec~for comparison analysis. HD~48682's disc was resolved by JCMT/SCUBA-2 observations at 450 and 850 $\micron$, where the images were also re-sampled to 1 pixel to 1\arcsec \citep{2017Holland}. We also located six sets of archival SMA observations of HD~48682 taken at $\approx$~230 GHz (Project ID 2015B-s014, P.I. Macgregor). Four out of the six SMA archival observation sets contained useable data.

\subsubsection{\textit{Herschel} Stellar position}

The optical position of HD~48682 is $6^h46^m46\fs44 \ {+33}\degr34\arcmin40\farcs97$  for the 100 $\micron$ source utilising the proper motions from \cite{2018G2}.  A 2D Gaussian profile was fitted to HD~48682, allowing for rotation and ellipticity. The centre of the disc profile was measured to within $1\farcs0$ from the optical position, which is well inside the \textit{Herschel} absolute pointing accuracy of $2\farcs5$ at the {1-$\sigma$} level \citep{2014ESan}.  For the deconvolution process presented here, we have assumed that the star's position is the centre of the 2D Gaussian profile.

\subsubsection{\textit{Herschel} Radial profiles and deconvolution}
\label{RD:radprof}

The extent of the {70~$\micron$} source was determined by fitting the semi-major and -minor axes of an ellipse; whilst the extent of the semi-major and -minor axes of the 100 and {160~$\micron$} maps were derived from the FWHM of model 2D Gaussian profiles along the major and minor axes.  The orientation (position angle, inclination) of the disc was consistent across all three images.  The disc radial extent is likewise consistent, after accounting for the larger PSF/FWHM at longer wavelengths, suggesting the disc is well-resolved by \textit{Herschel}.

To determine stellar contribution and to reveal debris disc structure, a reference point spread function (PSF) was used.  We adopted a similar technique applied by \cite{2017Hengst}, similar to the methods used in \cite{2010Liseau} and \cite{2011Marshall}.  Observations of a fiducial star, $\alpha$ B{\"o}otis (HD 124897, Arcturus), was selected, reduced and rotated to the same position angle observed for HD~48682, which was used as the instrument PSF.  This PSF was scaled to expected stellar photospheric flux corresponding to all three \textit{Herschel}/PACS wavebands, to then be centred on HD~48682's position and subtracted away from the respective original \textit{Herschel}/PACS maps.  This image result was then deconvolved with the instrument PSF by the Lucy-Richardson method \citep{1972Richardson,1974Lucy}.  See bottom row of Figure \ref{fig:stamps} for the deconvolved images.

\begin{figure}[ht]
\includegraphics[width=1\columnwidth, trim = 5.0cm 4.5cm 0cm 1.8cm]{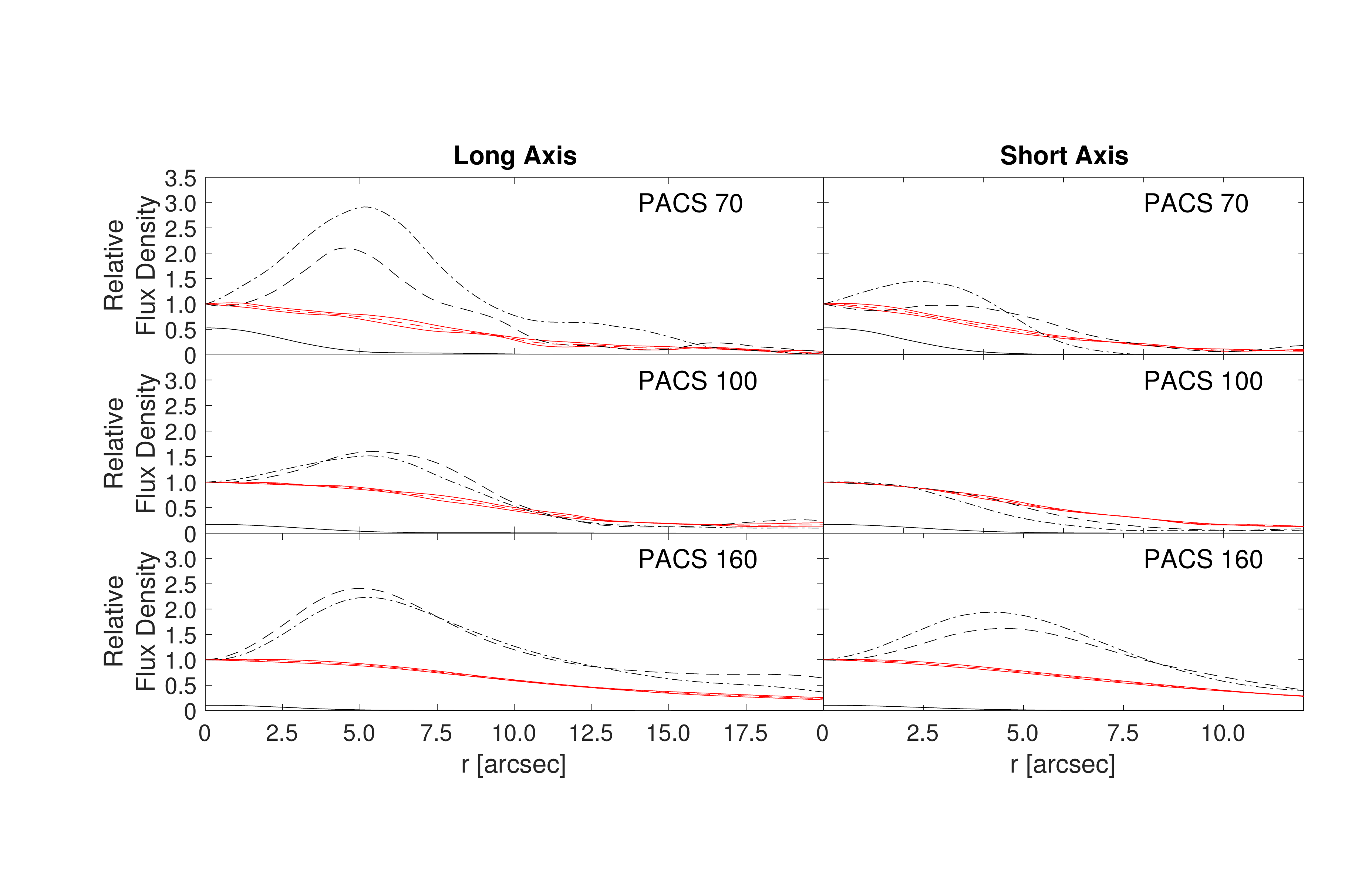}
 \caption{Radial profiles showing relative flux density as a function of arcsec for HD~48682 along the semi-major (left) and semi-minor (right) axes of the disc. The (red) dashed lines are the measured profile either side of the disc centre along each axis; the (red) solid errors bars mark the mean profile between each axis.  The (black) dashed lines and the (black) dashed-dotted lines are the deconvolved disc profiles along the NW arm and SE arm respectively for the long axis; along the NE extent and SW extent respectively for the short axis.  The deconvolved radial profiles have all been scaled to start at a relative value of 1.0 per arcsec for reference. The solid black line is the subtracted stellar profile, which has not been scaled for reference between between each map.}
 \label{fig:radial}
\end{figure}

After deconvolution, the semi-major and -minor axes of the disc were measured by fitting an ellipse to the region of the map, centred on HD~48682, which exceeded a 3-$\sigma$ threshold. The semi-major and -minor axes measured in all deconvolved images were relatively consistent, although slightly larger in in the 70 $\micron$ deconvolution. This could be attributed to the 3-$\sigma$ ellipse fit capturing what appears to be an extension branches of emission in the 70 $\micron$ map just beyond both the SE and NW branches of HD~48682 disc.  The distances from the star centre to the centre of the rings (both along NW and SE branches) were measured by fitting an ellipse to the (image) clump.  The coordinates of the centre of the ring was taken as the centre of the fitted ellipse.  Radial profiles are presented in Figure \ref{fig:radial} and a summary of corresponding measurements is presented in Table \ref{tab:rp}.

\subsubsection{Additional sources in the 70 $\micron$ map}  \label{bgs}
The 70 $\micron$ deconvolution appears to have revealed unknown sources beyond both NW and SE ansae (see Figure \ref{fig:radial}) away from the centre of HD~48682, which can be seen in Figure \ref{fig:stamps} as either an extension of the disc, background contamination, or random noise peaks. If these unknown sources are either background sources or noise peaks (i.e. confusion noise), then the corresponding flux estimate of 70 $\micron$ band would now be contaminated due to the radius of the photometric aperture used (12\arcsec).  Therefore, increasing the uncertainty in flux density measured in the aperture photometry conducted for the 70 $\micron$ source.

\subsubsection{Residuals}
\begin{figure*}
\centering
\includegraphics[width=2\columnwidth,trim = 5cm 2cm 5cm 2cm]{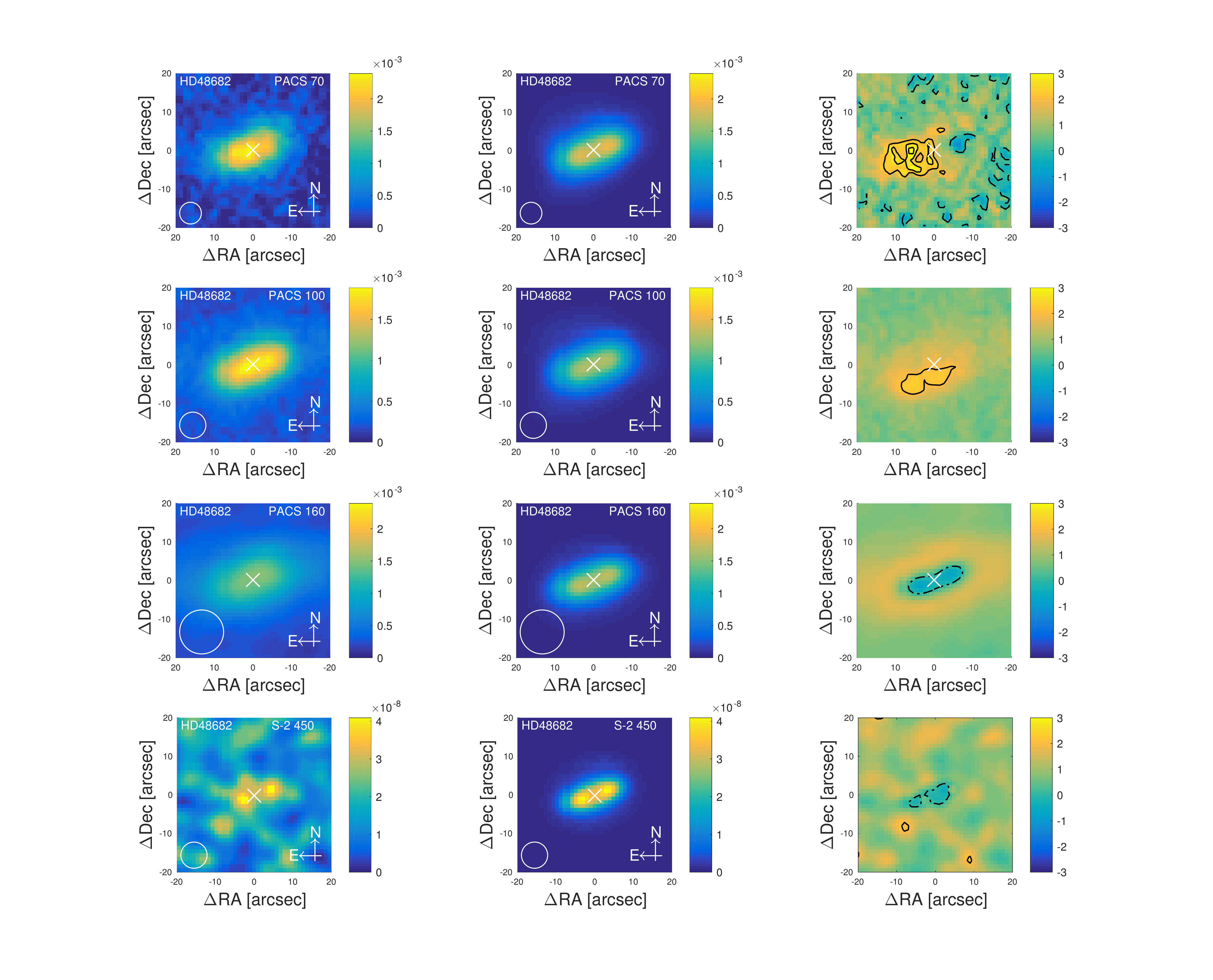}
\caption{Images of HD~48682 indicating potential asymmetry in 70 and 100 $\micron$ bands, further extension of emission in the 160 $\micron$ band, and SCUBA-2 450 $\micron$ for comparison. \textit{Left}: The original image with stellar subtraction. \textit{Middle}: The convolved model.  \textit{Right}: shows the residuals normalised by the RMS. Solid contours represent 2.0, 2.5, 3.0-$\sigma$ level where available, and the dot-dashed contours show the 0.0-$\sigma$ level where available to confirm disc subtraction. Image orientation is north up, east left. The images have the same scale of $1\arcsec$ pixel$^{-1}$.  The colour scale bar is in units of Jy/arcsec$^{2}$ for Columns 1 and 2. The colour bar scale for Column 3 are scaled to the respective RMS (in Jy arcsec$^{-2}$) values in each residual image, which are as follows: $2.3\times10^{-4}$ at 70 $\micron$, $2.6\times10^{-4}$ at 100 $\micron$, $3.6\times10^{-4}$ at 160 $\micron$, and $1.4\times10^{-8}$ at 450 $\micron$. 
\label{fig:asym}}
\end{figure*}

To determine any discernible architecture of HD~48682, models of the disc at each \textit{Herschel}/PACS and SCUBA-2 450 $\micron$ waveband were created using the fitted three-dimensional dust continuum model that was used in the radiative transfer code (see Section \ref{s:rt}), along with the disc orientation (111$\degr$) and inclination (58$\degr$) values determined from deconvolution of the 100 $\micron$ source. These models were further scaled to the expected flux density of the disc corresponding to their wavelength. These models were then subtracted from their respective original maps after removal of the stellar component by subtraction of a scaled PSF centred on the stellar location. The original images after stellar subtraction, the disc models, and residuals for all four wavebands are presented in Figure \ref{fig:asym}. We note here that the SCUBA-2 images provided by \cite{2017Holland} are available online\footnote{SONS Legacy Archive:\\https://www.canfar.net/citation/landing?doi=17.0005} in `png' format.  The `original' image corresponding to the SCUBA-2 450 $\micron$ waveband was created by reading in the pixel numbers from the `png' image into a 2D image array. A synthetic PSF was constructed by using the 2D Gaussian model described in \cite{2013Dempsey} as a guide to create the corresponding convolved disc model at 450~$\micron$.

Both the 70 and 100 $\micron$ sources show a brightness asymmetry with the stronger positive signal along the SE arm, with up to a 3-$\sigma$ detection for the 100 $\micron$ source. This can also be confirmed by the radial profiles. Asymmetry in the 70 $\micron$ source could be attributed to the background sources that were detected and/or more small grains being stirred.  There was no noticeable asymmetry detected in the 160 $\micron$ image, probably due to lower signal-to-noise of the disc and the larger beam FWHM, but rather we see a faint halo of residual extended emission surrounding the system. This might imply the disc model used is too compact, such that we are tracing a different grain population at 160 vs. 100~$\micron$ which has a more extended spatial distribution. However, the halo was not detected with any statistical significance and therefore we are not confident that this is a real component of the debris disc architecture. For the SCUBA-2 450 $\micron$ image the model appeared to adequately replicate the disc, highlighting a 2-$\sigma$ source near the SE ansa after model subtraction.

 \begin{table}
 \caption{Measurements of HD~48682's disc extent, orientation, and inclination. \label{tab:rp} }
 \centering
 \begin{tabular}{llll}
  \hline
 & 70$\micron$  &   100$\micron$     & 160$\micron$  \\
  \hline
  \hline
Original PACS Maps & & &\\
Semi-major [\arcsec] & 9.7 $\pm$ 0.9 & 11.3 $\pm$ 0.9 & 13.5 $\pm$ 2.3\\
Semi-major [au] & 162  $\pm$ 15 & 189.1 $\pm$ 15 &  226.9 $\pm$ 38 \\
Semi-minor [\arcsec] & 5.6 $\pm$ 0.8 & 6.4 $\pm$ 0.7 & 7.5 $\pm$ 1.7\\
Semi-minor [au] & 93 $\pm$ 13 & 107.0 $\pm$ 12 &  125.2 $\pm$ 28 \\
Inclination Angle [\degr] & 54.9 $\pm$ 5.2 & 55.6 $\pm$ 5.1  & 56.2 $\pm$ 8.3 \\
Position Angle [\degr] & 111.1 $\pm$ 0.5 & 110.4 $\pm$ 0.5 & 102.8 $\pm$ 1.0\\
\hline
Deconvolved Images & & & \\
Semi-major [\arcsec] & 13.2 $\pm$ 0.5 & 13.0 $\pm$ 0.5 &  12.8 $\pm$ 1.0 \\
Semi-major [au] & 221.5 $\pm$ 8.4 &  218.0 $\pm$ 8.4 & 214.9 $\pm$ 16.4\\
Semi-minor [\arcsec] & 8.2 $\pm$ 0.5 & 6.9 $\pm$  0.5 &  7.2 $\pm$ 1.0 \\
Semi-minor [au] & 137.0 $\pm$ 8.4 & 115.4 $\pm$  8.4 & 120.1 $\pm$ 16.4 \\
Inclination Angle [\degr] & 51.8 $\pm$ 6.9 & 58.0 $\pm$ 2.2  & 56.0 $\pm$ 7.7 \\
Position Angle [\degr] & 110.8 $\pm$ 0.5 & 111.4 $\pm$ 0.5 & 120.0 $\pm$ 1.0\\
NW Arm [\arcsec]  & 4.7 $\pm$ 0.5  & 5.6 $\pm$ 0.5 & 5.3 $\pm$ 1.0 \\
NW Arm [au] & 77.9 $\pm$ 8.0 & 93.1 $\pm$ 8.0 &  88.4 $\pm$ 17 \\
SE Arm [\arcsec]  & 4.8 $\pm$ 0.5 & 5.1 $\pm$ 0.5 & 4.7 $\pm$ 1.0 \\
SE Arm [au] & 79.8 $\pm$ 8.0 & 84.8 $\pm$ 8.0 &  79.1 $\pm$ 17\\
  \hline
 \end{tabular}
 \newline
 \end{table}

\subsection{Disc architecture}\label{s:DiscArc}

We assume HD~46682 has only a cold-component debris belt because the SED revealed no significant excess emission at mid-infrared wavelengths $\leq 20~\micron$, as can be seen in Figure \ref{fig:SED}. We then fitted a single (temperature) component, modified blackbody to the photometry at wavelengths where significant excess was measured ($>$ 3-$\sigma$), to calculate observational properties of the dust temperature, disc extent, and fractional luminosity ($T_{\rm dust}$, $R_{\rm dust}$, $L_{\rm dust}/L_{\star}$). A least-squares fit to the photometry weighted by the uncertainties calculated a fractional luminosity of $L_{\rm dust}/L_{\star}\,={\,(7.2~\pm~0.4)~\times10^{-5}}$ and temperature of {$T_{\rm dust}\,=\,57.9~\pm~0.7$~K}, in line with previous estimates of the disc brightness by \cite{2013Eiroa}. 

If we assume the dust acts like a blackbody source then we can estimate its radial location, which is referred to as the blackbody radius of the debris disc. The measured blackbody temperature of {HD~48682's} disc is equivalent to an orbital radius of {$R_{\rm dust}\,=\,31.2~\pm~1.1$~au}. A comparison with the mean distances of the NW and SE branches in the 70 and 100~$\micron$ deconvolutions ({78.9~$\pm$~11.3 au}) and ({89.0~$\pm$~11.3 au}) respectively, reveals that the ratio of the actual to blackbody radii to be $\Gamma\,=\, 2.7 \pm 0.5$. This is lower than expected ($\Gamma = 6.2 \pm 1.1$) when assuming pure astro-silicate grains ($\rho_s~=~3.3~g/cm^3$) determined from the models of \cite{2015PawKri}.

We determine the disc architecture through image analysis by initially assuming it is well approximated by a single Gaussian annulus, described by five parameters -- flux density $F_{\rm disc}$, radius $R_{\rm disc}$, fractional width $\Delta R_{\rm disc}$, inclination $\theta$, and position angle $\phi$. Here we have presented results based on fitting the model to the 100~$\mu$m image data alone. We note that similar values for the disc architecture are obtained by fitting the 70~$\mu$m data, but with greater uncertainties due to the lower signal-to-noise of the observation at that wavelength.

\begin{figure*}
\centering
\includegraphics[width=\textwidth]{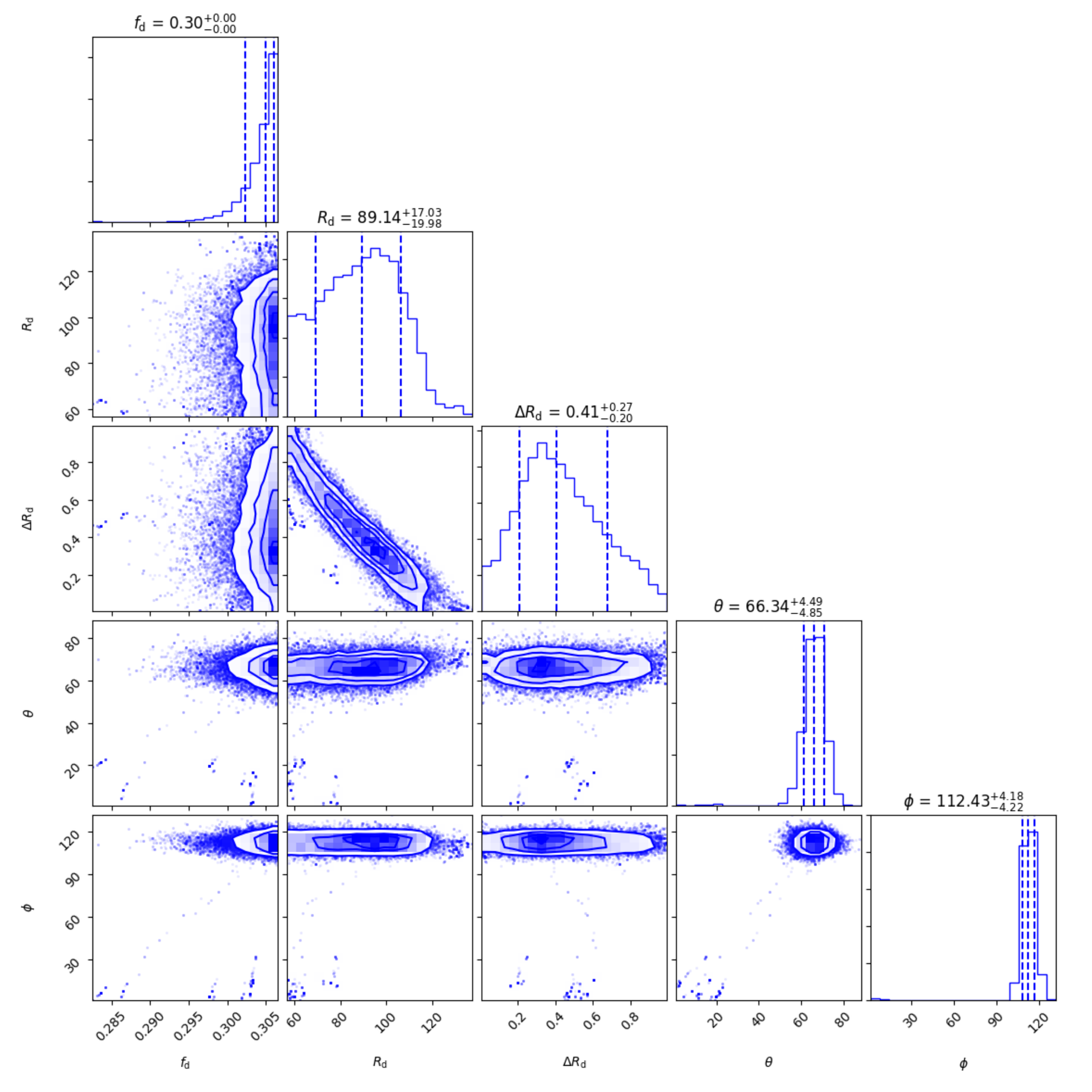}
\caption{Corner plot illustrating the outcome our MCMC analysis carried out with {\sc emcee}. A single Gaussian annulus was used to model the disc, and convolved with a PSF before comparison with the \textit{Herschel}/PACS 100~$\mu$m images. In total 125,000 realisations (250 walkers, 500 steps) were made, with the first 200 steps of each chain discarded as burn-in, see text for further details.
\label{fig:emcee_output}}
\end{figure*}

To determine the maximum likelihood values for each model we adopt a Bayesian approach using the affine invariate Markov chain Monte Carlo (MCMC) code \textsc{emcee} \citep{2013Foreman-Mackey} to explore the parameter space of our discs, and determine the best-fit parameters (maximum probability) for each disc and their associated uncertainties (16$^{\rm th}$ and 84$^{\rm th}$ percentiles of the probability distribution). This method has the advantages of being relatively fast to comprehensively explore parameter space before converging on an optimum solution, and dealing well with degeneracies between disc semi-major axis and inclination that we expect to see in the case of marginally resolved systems such as are represented in the data used here.

The initial values for the model disc parameters were determined from the measured flux density and a 2D-Gaussian fit to the source brightness profile, except for the fractional width of the disc ($\Delta R_{\rm disc}$), which was initially set as 0.2. The distribution of initial conditions were likewise based on observed properties of the disc, or instrumental limits. The instrument calibration uncertainty for the SCUBA-2 450~$\micron$ was calculated to be approximately 12\% \citep{2013Dempsey}. The disc flux density of HD~48682 at 450~$\micron$ was measured to be about 32~mJy/beam, with an image RMS of 4.7~mJy/beam \citep{2017Holland}.  Therefore, the disc flux density was constrained to lie within $\pm$~20\% of the measured value. The disc radius was constrained to lie between an extent equivalent to half the beam FWHM (i.e. the source must be spatially resolved) and twice the measured FWHM from the 2D Gaussian fit. The fractional disc width was given free range between 0.1 and 0.9. The disc position angle (0--180\degr) and inclination (0--90\degr) were both constrained to lie within their respective ranges. 

To construct an objective function, we sum the residuals (observation - convolved model) of each realisation for all pixels within a mask. The mask is defined in the following manner. We first identify pixels in the observed image with flux density values greater than 3-$\sigma$ lying within a circular region of radius 3$\times$FWHM centred on the star. This initial mask is then convolved with the instrument PSF to extend it to adjacent regions to avoid over-fitting the model to the central regions of the source brightness distribution. The final mask area thus comprises all pixels of the convolved mask with values $>$ 0.1.

We ran a total of 125,000 realisations of the model with 250 walkers and 500 steps. We used the first 200 steps as a burn-in for the MCMC chains and calculated the probability distributions from the final 75,000 realisations. From this analysis we determine the maximum likelihood disc parameters to be $F_{\rm disc} = 305^{+1}_{-3}$ mJy, $R_{\rm disc} = 89^{+17}_{-20}~$au, $\Delta R_{\rm disc} = 0.41^{+0.27}_{-0.20}$, $\theta = 66\fdg7^{+4.5}_{-4.9}$, $\phi = 112\fdg5^{+4.2}_{-4.2}$. These values are consistent, within uncertainties, with estimates previously published in the literature based on the same data set \citep[see ][]{2013Eiroa}. The {\sc{emcee}} disc values are presented in Table \ref{tab:pl} to compare with the {\sc{Hyperion}} RT values (see Section \ref{s:rt}).

The inferred disc radius from this work is consistent with the predicted value ($R_{\rm disc} = 81.2^{+8.7}_{-8.3}~$au) for a star with the same luminosity as HD~48268 ($L_{\star} = 1.752~L_{\odot}$) based on the relation determined in \cite{2018Matra} from a sample of millimetre-resolved debris discs. This consistency is satisfying, given that even at far-infrared wavelengths the peak of dust emission from a disc should still roughly trace the planetesimal belt location in a system \citep{2019Pawellek}. Since the disc is bright for its age \citep[$L_{disc}/L_{\star}$~7.2$\times10^{-5}$, $>1$Gyr;][]{2011Kains}, we might infer that the broad disc is indicative of some stirring.

\subsection{Radiative transfer}\label{s:rt}

We use the radiative transfer (RT) Monte-Carlo code, {\sc{Hyperion}} \citep{2011Robitaille}, to simultaneously fit the extended emission and SED of HD~48682.  {\sc{Hyperion}} allows the user input a three-dimensional dust continuum model, setting this apart from the typical power-law radial distribution models.  

\begin{eqnarray}
D_g(r,h) &=& \exp{\left[-\frac{r-r_m}{\sqrt{2}r_w}\right]^2}f(r,h) \label{eq:guassian} \\
f(r,h) &=& \frac{1}{2}\exp{\left[-\frac{|h|}{h_{s}r}\right]^2} \label{eq:sup}
\end{eqnarray}

Equation \ref{eq:guassian} shows the distribution of the density of dust grains: where $r$ is the radial position,  $r_m$ is the mean distance, $r_w$ is the width component of the Gaussian, $h$ is the height above or below the midplane, and $h_s$ is the scale height which can modify the thickness of the debris disc in equation \ref{eq:sup}. For simplicity in this model, $h_s$ is kept constant at 0.1. We believe the Gaussian belt model used here is adequate to represent the dust distribution, and facilitates comparison with modelling work on similar systems.  We assumed a dust composition of astronomical silicate with a density of  $\rho~=~3.3~g/cm^3$ \citep{2003Draine}.

From the 100 $\micron$ deconvolution, the inner and outer edge of the disc were determined to be approximately 28 au and 111 au respectively and from the MCMC fitting, the radius of the disc to be 89.1$^{+17.0}_{-20.0}$~au. So, for the purpose of modelling, $r_m$ was allowed to vary between 60 au and 100 au, whilst the $r_w$ was allowed to vary between 25 au and 65 au.  The constituent grains were represented by a power law size distribution (s$^{-q}$ds) between $s_{\rm min}$, which was allowed to vary between 0.5 and 5.0 $\micron$, and $s_{\rm max}$, which was fixed at 3000 $\micron$, with the exponent, $q$, allowed to vary from 3.0 to 4.0.

The {\sc{Hyperion}} RT code with the Gaussian debris belt model managed to fit the SED data, specifically fitting to the rising \textit{Spitzer} IRS spectrum and the falling of the sub-mm data with a $\chi^2_{red}$ of 1.28 (13 degrees of freedom); see Figure \ref{fig:SED} for the SED fit.  The mean distance of the disc lies at $r_m$ of $82^{+3}_{-2}~{\rm{au}}$ with a $r_{w}$ component of {$42^{+2}_{-1}$~au}. The minimum dust grain size was calculated to be 0.60$^{+0.06}_{-0.07}~$ $\micron$, with an exponent of the grain size distribution to be $q = 3.60 ^{+0.02}_{-0.01}$. The uncertainty values for the {\sc {Hyperion}} calculations are smaller than the MCMC deviations can be attributed to modelling photometric values as oppose to image data. See Table \ref{tab:pl} for the summary of the variables in the parameter space and the fitting results for the {\sc{Hyperion}} RT code with the disc parameters from the MCMC analysis are presented for comparison. 

\begin{table}
\caption{{\sc{Hyperion}} RT code parameter space results compared with {\sc{emcee}} code results.  Note $\nu$ (= N - f) is the number of degrees of freedom. Where $N$ is the number of data points and $f$ is the number of variables used in the parameter space. \label{tab:pl} }
\centering
\begin{tabular}{lllll}
 \hline
Parameter &  Range  &   Distribution   & {\sc{Hyperion}} & {\sc{emcee}} \\
 \hline\hline
$r_{\rm m}$ (au) & 60 -- 100 & Linear & 82$^{+3}_{-2}$ & 89.1$^{+17.0}_{-20.0}$\\
$r_{\rm w}$ (au) & 25 -- 65 & Linear & 42$^{+2}_{-1}$ & 36.5$^{+24.1}_{-17.8}$\\
$s_{\rm min}$ ($\micron$) & 0.5 -- 5.0 & Linear & 0.60$^{+0.06}_{-0.07}$ &\\
$s_{\rm max}$ ($\micron$) & --- & Fixed & 3000 & \\
$q$ & 3.00 -- 4.00 & Linear & 3.60$^{+0.02}_{-0.01}$  & \\
Composition & astron. sil. & Fixed & ---& \\
$\chi^{2}_{red},~\nu$ & --- & --- & 1.28, 13 &\\
 \hline
\end{tabular}
\newline
\end{table}

\subsection{Sub-millimetre observations}

HD~48682's disc was resolved by JCMT/SCUBA-2 observations at 450 and 850 $\micron$ tracing larger, cooler dust grains. The analysis of the JCMT/SCUBA-2 {450~$\micron$} image (RMS = 4.7 mJy/beam) by \cite{2017Holland} measured a couple of (marginal) 3-$\sigma$ peaks of {16.0~mJy} (SE) and {16.7~mJy} (NW) around the HD~48682's stellar position of the disc at its ansae.  The {450~$\micron$} source may reveal a position asymmetry where the location of the SE and NW ansae are measured at {71.6~$\pm$~12.6~au} and {97.7~$\pm$~14.5~au} respectively. The 850 $\micron$ image indicates a brightness asymmetry where the NW peak is brighter.

\begin{figure*}
\centering
\includegraphics[width=2\columnwidth,trim = 0cm 0cm 0cm 0cm]{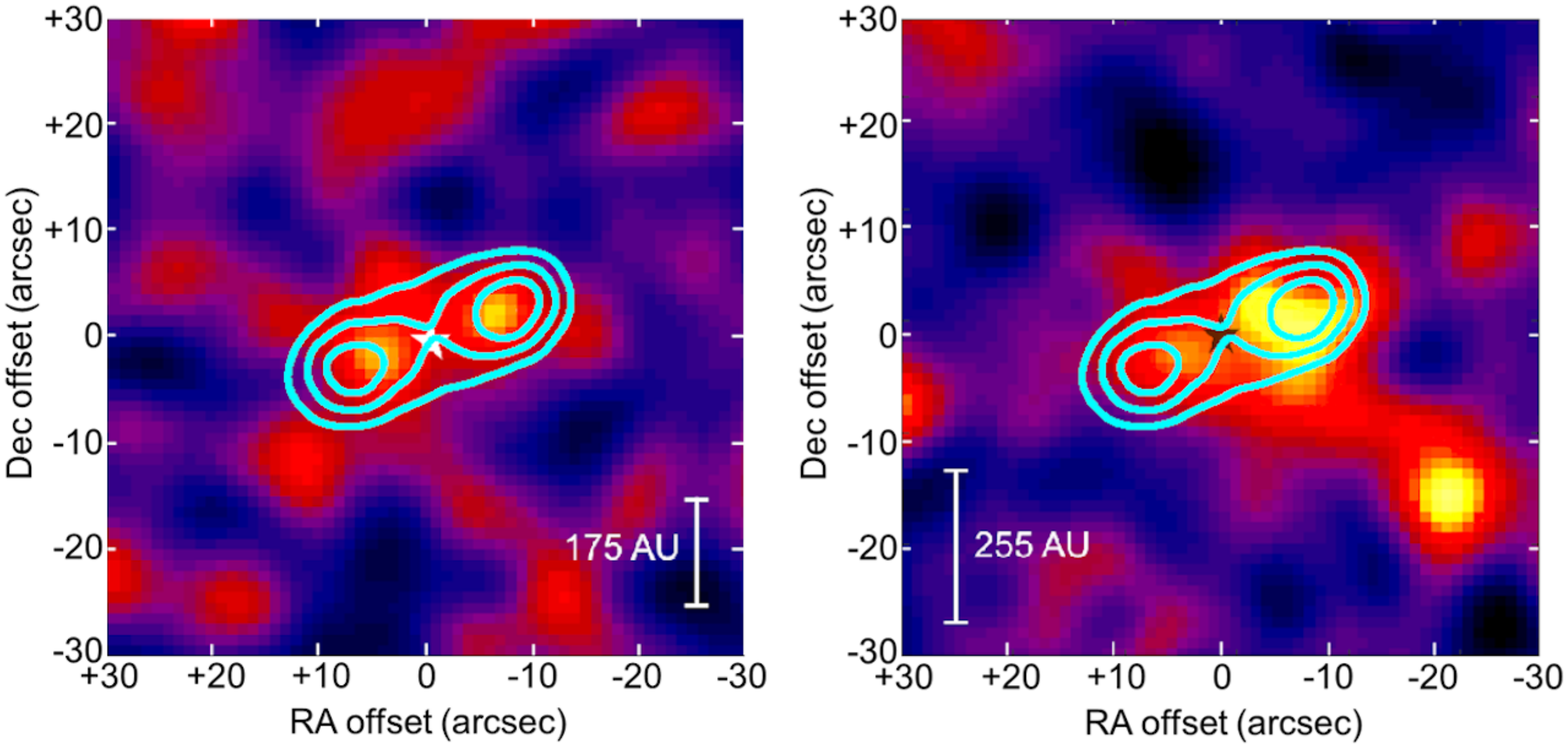}
\caption{JCMT/SCUBA-2 450 $\micron$ (left) and 850 $\micron$ (right) image of HD~48682 revealing the disc architecture at millimetre wavelengths \citep{2017Holland}. The star symbol denotes the optical position of HD~48682. The contour lines show the 2.0, 3.0, and 4.0-$\sigma$ significance levels of the deconvoluted 100 $\micron$ image. Images are taken from SONS: Legacy dataset available online \citep[doi:10.11570/17.0005;][]{2017Holland}.}
\label{fig:Scuba450850}
\end{figure*}

We calibrated and reduced the usable SMA observations using the {\sc Miriad} software package following the data reduction scripts on the SMA webpage. Uranus was used as the flux calibrator, 0646+448 and 0555+398 were phase calibrators, and 3c84 and 3c273 were used as the bandpass calibrators. The final reduced continuum image was centred on the stellar position with an RMS noise level of 0.2 mJy/beam and a beam FWHM of 4\farcs25$\times$4\farcs02. As expected, there was no detection at the stellar position of HD~48682 in the map, but we also find no convincing evidence for detection of the disc in the observation (assuming a predicted flux density of 1.9~mJy, extrapolated from the 850~$\mu$m measurement). The level of noise in the image is too high for us to put any constraints on the disc width or radius, as all plausible architectures from our modelling lie well below the threshold for detection. We can however state that images at least an order of magnitude deeper (10 to 20 $\mu$Jy/beam) are required to detect the disc with any confidence, putting this target beyond the reach of most Northern hemisphere millimetre-wavelength facilities. However, there was a marginal, ~3-$\sigma$ (0.6 mJy), detection consistent with the peak position close to the western side of the disc in the JCMT/SCUBA-2 850 $\micron$ image (see Figure \ref{fig:Scuba450850}). We speculate from this that much, if not all, of the asymmetry exhibited in the 850~$\mu$m image can be explained as the result of background contamination from this extended emission. 

\begin{figure}
\centering
\includegraphics[width=\columnwidth,trim = 1cm 0cm 1cm 0cm]{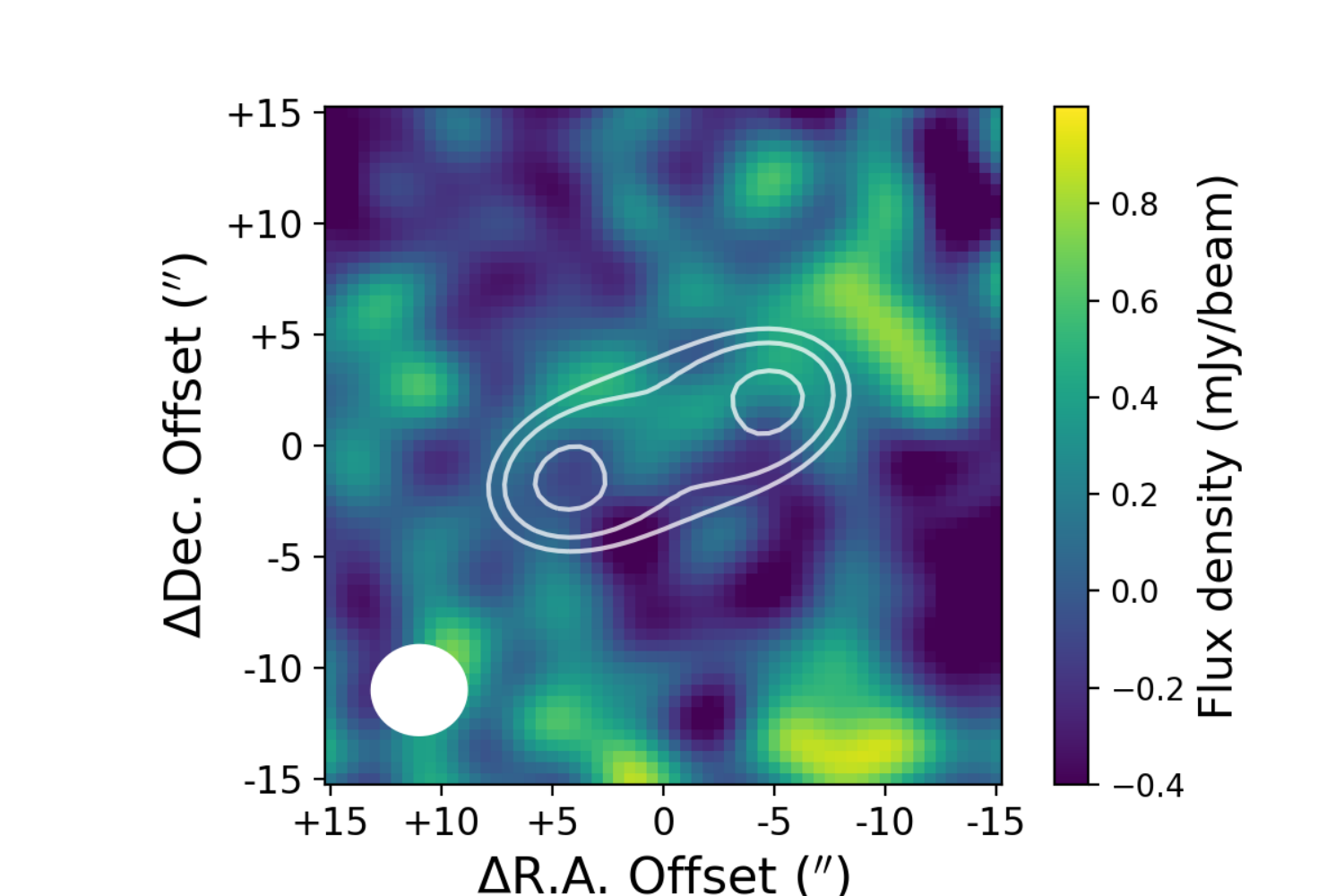}
\caption{SMA 225 GHz observation of HD~48682. The image colour scale is linear between -0.4 to 1.0 mJy/beam, equivalent to -2 to 5-$\sigma$. The white contours denote the expected extent and orientation of the debris disc, based on the maximum probability model from the MCMC fitting ($R_{\rm peak} = 89~$au, $\Delta R = 0.41$); the contours denote 0.5, 0.68, 0.9$\times$ the peak flux of the model ($\sim$200~$\mu$Jy/beam -- note that the disc is too faint to be detectable in the observation. The image orientation is North up, East left, the plate scale is 0.5\arcsec/pixel, and the instrument beam is denoted by the white ellipse in the bottom left corner of the plot ($\sim 4\arcsec$ FWHM).}
\label{fig:SMA225}
\end{figure}

\section{Discussion}\label{s:D}

Here we place the results of our analysis into context, and examine the origins and the characteristics of the disc around HD~48682. 

\subsection{State of the disc}

From the deconvolution of the \textit{Herschel}/PACS 100~$\micron$, we take the observed radius of the disc to be 89~$\pm$~11~au, calculated as the mean distance of the two ansae about stellar position.  The maximum likelihood probability from the MCMC analysis determined the fractional width is greater than the Solar system's Edgeworth-Kuiper belt, but not as broad as those discs identified as having substructure due to disc-planet interactions, i.e. HD 107146 \citep{2017Su} and HD 95086 \citep{2018Marino}.

The ratio of observed radius to the blackbody radius of the disc ($\Gamma = 2.9 \pm 0.5$) is smaller than expected if assuming pure astro-silicate grains ($\Gamma = 6.2 \pm 1.1$) when correlated with stellar luminosity \citep{2015PawKri}. On the other hand, if we assume the grains are composed of 50\% astro-silicate grains and 50\% ice ($\rho_s = 2.3 g/cm^3$, `icy grains'), the ratio is closer to what is expected ($\Gamma = 4.0 \pm 0.5$). However, if we consider the upper limit of the radius and fractional width results of the MCMC analysis, we calculate an outer edge of the disc to be at {$\approx$~142 au}, i.e. considering a disc almost twice its size. Then this distance corresponds to $\Gamma \approx 4.5$, which is getting closer to the expected value when assuming pure astro-silicate grains.

The disc model we assumed to determine the dust properties using {\sc{Hyperion}}, based upon a Gaussian annulus, calculated that HD~48682 has a disc of radius 82$^{+3}_{-2}$ au and a width of 42$^{+2}_{-1}$ au within uncertainties of the measured disc extent of \textit{Herschel}/PACS 100~$\micron$ deconvolution.  The disc was found to be dominated by small grains, assuming a grain composition of astro-silicate, with a size distribution exponent of 3.60 $\pm~0.02$ and a minimum grain size of {0.6~$\pm$~0.1~$\micron$} when integrated up to a grain size of 3 mm. The size distribution exponent is steep by comparison to previous studies of millimetre detected discs of \cite{2016Macgregor}, $q~=~3.35~\pm~0.02$, and \cite{2017Marshall}, $q~=~3.15~\pm~0.09$ and slightly steeper than the often assumed steady state infinite collisional cascade  \citep[$q = 3.5$,][]{1969Dohnanyi}. The minimum grain size is lower than expected when compared to similar systems studied by \citep{2014Pawellek} and 3-4 times smaller when compared with the blowout size for collisional active discs \citep[$3<q<4$,][]{2018AHughes}. Where the grains are acting under the assumption of compact spherical grains of astronomical silicate \citep{2003Draine}. 

However, all grains in the HD~48682's disc (located from $\sim$~40~au to $\sim$110~au) could have a composition of dust silicate and water ice since the sublimation distance of these grains is about 27~au \citep{2011Kobayashi}, assuming a sublimation temperature of 100~K for icy grains. Some of this icy contribution to the dust could be expected given the results from similar analyses of \textit{Herschel}-resolved discs. This suggests the presence of moderately icy grains \citep{2016Morales} and the increasing number of CO detections in debris discs \citep[e.g.][]{2013Moor,2016Greaves, 2016Marino, 2017Marino, 2017Matra} revealing the planetesimals in exo-Kuiper belt analogues to be volatile rich \citep[although detection of CO in a G-star debris disc is not expected; e.g.,][]{2017Kral,2020Marino}.

HD~48682 is also unusual where the disc is smaller than systems with similar age \citep{2011Kains,2014Pawellek}, albeit still considered a broad system with a fractional width of $\sim$0.5. Debris discs of age greater ($> 1$~Gyr) typically have a radial extent of 180~$\pm$~45~au, where the disc of HD~48682 is about half this size. HD~48682's debris disc is bright in the thermal emission for its age \citep[$> 1$~Gyr,][]{2011Kains} when compared to the trends that are expected for Sun-like stars \citep[e.g.][]{2001Spangler,2003Decin,2018Sibthorpe}. Collionsally active discs produce smaller grains that overheat due to inefficient absorption and scattering properties \citep{1993Backman}.  Therefore, the compact nature of the disc and the small grains may attribute to a brighter nature of HD~48682's disc.

\subsection{Potential asymmetry and halo}

The deconvolved images revealed a potential asymmetry in the separation and brightness of the two ansae from the stellar position.  The asymmetry appeared to be verified when the structure and emission from the disc around HD~48682 was not effectively replicated with an axisymmetric disc model for the 70 and 100 $\micron$ sources.  However, the disc model was found to replicate the 160 $\micron$ source. In comparison with the JCMT/SCUBA-2 {850~$\micron$} image \citep{2017Holland}, it shows a brightness asymmetry but the brighter ansa is on the NW side of the disc, which is on the opposite side to what has been in observed in the far-infrared 70 and {100~$\micron$} sources. Figure \ref{fig:Scuba450850} (right image) shows the differences in brightness asymmetry by presenting the {850~$\micron$} SCUBA-2 image with a contour overlay of the \textit{Herschel}/PACS {70~$\micron$} deconvolution. There is also a 2-$\sigma$ detection in the {SMA-225~GHz} image that is NW of HD~48682's position (see Figure \ref{fig:SMA225}).

The structure being visible in both the JCMT and SMA images could be due to the emission originating from a brightness asymmetry within the disc. If so, this may indicate that the grains may be spatially segregated into different orbital and (possibly) radial locations according to their size. If we assume the NW ansae in the SCUBA-2 850~$\micron$ image is real, then this may be an example of apocentre glow \citep[e.g.][]{2016Pan}.  Assuming the background galaxy contamination of 1.2~mJy/beam \citep{2017Holland} in the NW peak and little or no contamination in the SE peak, then rough approximations of the fluxes of apocentre and pericentre peaks are about $\sim$1.6~mJy/beam and $\sim$1.3~mJy/beam respectively. The eccentricity of the disc at 850~$\micron$ is estimated to be about 0.23 by using the apocentre/pericentre flux ratio to eccentricity relationship derived by \cite{2016Pan}.  This value is similar to eccentricity calculated from the 450~$\micron$ image of 0.19 using apocentre and pericentre distance values. 

The presence of non-axisymmetric structure (eccentricity, brightness asymmetry) in a disc often attributed to perturbation of the disc by an unseen companion. As a relatively old system ($>$~1~Gyr), a brightness asymmetry in HD~48682's disc would be of intense interest regarding the stirring of planetesimal belts at late times, such as the proposed Late Heavy Bombardment in the Solar system \citep{1974Tera,2018Lowe}, or the bright disc around HD 10647 \citep{2016Schuppler}. However, the brightness asymmetry is potentially due to contamination (from e.g. Galactic cirrus emission or a background galaxy). Therefore, we cannot draw any definitive conclusions regarding the shape of the disc based on the available observations. 

\section{Conclusion}\label{s:C}

In this work, we have presented archival far-infrared (\textit{Herschel}/PACS) and sub-millimetre (JCMT/SCUBA-2) observations of the debris disc around HD~48682. These images have revealed emission from a disc extended along both its semi-major and semi-minor axes in all four wavebands. The disc extent was determined by MCMC analysis of the 100~$\micron$ image to be {89$^{+17}_{-20}$~au} with an inclination to be 66$\fdg3^{+4.5}_{-4.9}$ at a position angle of 112$\fdg4^{+4.2}_{-4.2}$.

The location of the cold debris belt surrounding {HD~48682} was measured to have a mean distance of {83.4~$\pm$~20.4~au} and {84.6~$\pm$~19.2~au} in the \textit{Herschel}/PACS and SCUBA-2 {450~$\micron$} images, respectively. This was verified by modelling of the disc using codes {\sc Hyperion} to model the disc and {\sc MCMC} to infer the maximum probability model for the disc and its uncertainties. The measured planetesimal belt radius is consistent with other systems of similar luminosity studied at sub-millimetre wavelengths \citep{2018Matra}. The deconvolved images and the modelling also revealed the disc to be (moderately) broad, from an inner edge of {$\sim$~40~au} to an outer edge of {$\sim$105~au}.

A marginal brightness asymmetry was observed with the {70~$\micron$} sources after deconvolution, where a greater flux density was measured for the ansa SE of {HD~48782}.  A brightness asymmetry was also observed in the {850~$\micron$} source but this time on the NW side of {HD~48682}.  This could suggest an pericentre glow \citep[e.g.][]{1999Wyatt} at far-infrared wavelengths and a apocentre glow \citep[e.g.][]{2016Pan} at sub-millimetre wavelengths, similar to what was revealed in the Fomalhault system \citep{2017MacGregor}, corresponding to a possible eccentricity of about $\sim$~0.2.  This could suggest dust segregation due to grain size perturbed by an unseen companion, however, we do caution that due to the low signal-to-noise of the data, this could be attributed to emission from background contamination.

Combining the disc SED with the multi-wavelength, spatially resolved images we fitted a 3D dust continuum model, assuming the disc to be a single annulus around the star, and calculated a minimum grain size of {0.6~$\pm~$0.1$~\micron$}, which is substantially smaller than compared with similar systems studied by \citep{2014Pawellek}. The sub-millimetre photometry has enabled us to constrain the grain size distribution with an exponent of $q$ = 3.60~$\pm$~0.02, thus, the debris disc has a grain size distribution that is slightly steeper than the oft-assumed infinite steady-state collisional cascade value \citep{1969Dohnanyi}.

HD~48682's debris disc is bright in continuum emission compared to systems of similar luminosity and age \citep{2011Kains}, which can be   attributed to the cold belt being both broad and dominated by smaller grains that are near the limit at which they are blown out of the system by radiation pressure. The presence of a dominant population of small dust grains in the disc could account for its breadth, with these small grains quickly driven onto eccentric orbits under the influence of radiation pressure \citep{2010Krivov}. Further imaging of the system in scatter light and millimetre wavelengths will be necessary to differentiate between the possible architectures of HD~48682's broad disc, either with sub-structure similar to HD~107146 \citep[e.g.][]{2018Marino}, or with an extended halo of dust grains similar to HD~61005 \citep[e.g.][]{2018Macgregor}.

\section{Data availability}

The datasets were derived from sources in the public domain:\\
\textit{Herschel Science Archive}:\\http://archives.esac.esa.int/hsa/whsa/\\
\textit{Hipparcos Input catalogue}:\\ https://www.cosmos.esa.int/web/hipparcos/input-catalogue\\
\textit{NASA/IPAC Infrared Science Archive}:\\ https://irsa.ipac.caltech.edu/\\
\textit{Gaia Archive}:\\ https://gea.esac.esa.int/archive/\\
\textit{Castelli \& Kurucz Atlas}:\\ https://archive.stsci.edu/hlsps/reference-atlases/cdbs/grid/ck04models/\\
\textit{JCMT Science Archive}: \\ https://www.eaobservatory.org/jcmt/science/archive/\\

\section*{Acknowledgements}
The authors would like to thank the reviewer, Dr~Jane~S.~Greaves, for her invaluable feedback, which has improved the quality and the conclusions of this paper. This research has made use of the SIMBAD database, operated at CDS, Strasbourg, France.  This research presented here in this paper has made use of the VizieR catalogue access tool, CDS, Strasbourg, France and NASA's Astrophysics Data System. JPM acknowledges research support by the Ministry of Science and Technology of Taiwan under grants MOST104-2628-M-001-004-MY3 and MOST107-2119-M-001-031-MY3, and Academia Sinica under grant AS-IA-106-M03.

\textit{Facilities:} \textit{Herschel}, JCMT, SMA, \textit{Spitzer}.

\textit{Software:} {\sc Hyperion} \citep{2011Robitaille}, {\sc{emcee}} \citep{2013Foreman-Mackey}, \textit{corner} \citep{2016Foreman-Mackey}, \textit{matplotlib} \citep{2007Hunter}.




\bibliographystyle{mnras}
\bibliography{papers} 


\bsp	

\label{lastpage}
\end{document}